\documentclass[12pt,a4paper]{article}
\usepackage{graphicx,epsfig}
\usepackage{amsfonts} 
\usepackage{amsthm, amsmath}
\usepackage{bbm}
\usepackage{indentfirst}  
\usepackage{listings}
\usepackage{caption}
\usepackage{subfig}
\usepackage{epstopdf}
\usepackage{stackengine}
\usepackage{multirow}
\usepackage{bbm,amsmath}
\usepackage{textcomp}
\usepackage{titling}
\usepackage{verbatim}
\usepackage{wrapfig}
\usepackage{float}
\usepackage{csquotes}
\usepackage{tabularx, booktabs}
\usepackage{array}
\newcolumntype{P}[1]{>{\centering\arraybackslash}p{#1}}
\usepackage{authblk}

\def\reals{\mathbb{R}}

\def\minimize{\mathop{\rm minimize}\limits}
\def\ovr{\mathop{\rm over}}
\def\st{\mathop{\rm subject\ to}}

\begin{document}

\title{Pricing index options by static hedging under finite liquidity}
\author{John Armstrong, Teemu Pennanen, Udomsak Rakwongwan\thanks{udomsak.rakwongwan@kcl.ac.uk}}
\affil{Department of Mathematics, \\King's College London, \\Strand, London, WC2R 2LS, United Kingdom}


\maketitle

\begin{abstract}
We develop a model for indifference pricing in derivatives markets where price quotes have bid-ask spreads and finite quantities. The model quantifies the dependence of the prices and hedging portfolios on an investor’s beliefs, risk preferences and financial position as well as on the price quotes. Computational techniques of convex optimisation allow for fast computation of the hedging portfolios and prices as well as sensitivities with respect to various model parameters. We illustrate the techniques by pricing and hedging of exotic derivatives on S\&P index using call and put options, forward contracts and cash as the hedging instruments. The optimized static hedges provide good approximations of the options payouts and the spreads between indifference selling and buying prices are quite narrow as compared with the spread between super- and subhedging prices.  
\end{abstract}

\section{Introduction}

In incomplete markets, the prices of financial products offered by an agent depend on subjective factors such as views on the future development of the underlying risk factors, risk preferences, the financial position as well as the trading expertise of the agent. A agent's prices also depend on the prices at which the agent can trade other financial products since that affects the costs of (partial) hedging when selling a product.

The {\em indifference pricing} principle provides a consistent way to incorporate the above factors into a pricing model. A classical reference on indifference pricing of contingent claims under transaction costs is \cite{hn89}. In the insurance sector, where market completeness would be quite an unrealistic assumption, indifference pricing seems to have longer history; see e.g.\ \cite{buh70}. A more recent account with further references can be found in \cite{car9}.

Indifference pricing builds on an optimal investment model that describes the relevant sector of financial markets as well as the agent's financial position, views and risk preferences. Realistic models are often difficult to solve much like the investment problem they describe. This paper develops a computational framework for indifference pricing of European style options on the S\&P500 index. Instead of the usual dynamic trading of the index and a cash-account, we take index options as the hedging instruments. For the ease of implementation, we consider only buy-and-hold strategies in the options but we take actual market quotes as the trading costs. For the nearest maturities, there are some $200$ strikes with fairly liquid quotes. This results in a convex stochastic optimization problem with one-dimensional uncertainty but over $400$ decision variables. The model is solved numerically using discretization and an interior point solver for convex optimization. The indifference prices for a given payout are found within seconds so it is easy to study the effect of an agent's views, risk preferences and financial position on the indifference prices. 

Much like the Breeden--Litzenberger formula, indifference pricing provides automatic calibration to quoted call option prices. While the Breeden--Litzenberger formula provides only a heuristic approximation in real markets with only a finite number of strikes and finite liquidity, the indifference approach finds the best static hedge given the quotes, the agents views and preferences. Moreover, the indifference approach gives explicit control of the hedging error in incomplete markets. Unlike the Breeden-Litzenberger formula would suggest, we find that in the presence of bid-ask spreads, the optimal hedges are often quite compressed portfolios of options taking positions only in few of the strikes. This is a significant benefit when implementing the hedges in practice. While the Breeden--Litzenberger formula applies only to options whose payouts are differences of convex functions of the underlying, the indifference pricing applies just as well to discontinuous payoffs such as digital options.

\section{The market}

We study contingent exchange traded claims with common maturity $T$ and payouts that only depend on the value of the S\&P500 index at $T$. This includes put and call options, forward contracts and cash. In general, lending and borrowing rates for cash are different, so the payout on cash depends nonlinearly on the position taken. Similarly, the forward rates available in the market depend on whether one takes a long or short position. For the options, on the other hand, the payout per unit held is independent of the position. The payoffs for holding $x\in\reals$ units of an asset are given in Table~\ref{table:payoffs}.

\begin{table}[h!]
\centering
  \resizebox{0.5\textwidth}{!}{  
\begin{tabular}{l|l} 
 \hline
 Asset & Payoff with $x$ units held\\ 
\hline
 Cash & $\min\{e^{r^aT}x,e^{r^bT}x\}$ \\ 
 Forward & $\min{\{(X_T-K^a)x,(X_T-K^b)x\}}$\\
 Call & $\max{\{(X_T-K),0\}x}$ \\
 Put & $\max{\{(K-X_T),0\}}x$\\
 \hline
\end{tabular}
}
\captionsetup{width=1\textwidth}
\caption{The payoffs of holding $x$ units of the assets. Here $r^a$ and $r^b$ and the borrowing and lending rates, respectively, $X_T$ is the value of the underlying at maturity, $K^a$ and $K^b$ are forward prices for long and short positions, respectively and $K$ is the strike price of an option}
\label{table:payoffs}
\end{table}

While the option payoffs are linear in the position, the cost of entering a position depends nonlinearly on the units $x$. For a long position $x>0$, one pays the ask-price while for short position, one gets the bid-price. The cost of buying $x$ units of cash is simply $x$ while for the forward, the cost is zero.

For each contract, the market quotes come with finite quantities. For the nearest maturity, one can find quotes for some 400 options on S\&P500. Table~\ref{table:BlooembergQuotes} gives an example of quotes available on the 8 April 2016 at 14:55:00 for contracts expiring on 17 June 2016.




\begin{table}[h!]
\centering
 \resizebox{0.9\textwidth}{!}{  
\begin{tabular}{l|c|c|c|c|c}
 \hline 
 Ticker & Type & Bid quantity & Bid price & Ask price & Ask quantity \\
 \hline\hline
 ESM6 Index & Forward & 258 & 2048.75 & 2049 & 377 \\
 SPX US 6/17/2016 C2095 Index & Call & 623 & 26.90 & 28.20 & 506 \\
 SPX US 6/17/2016 P2095 Index & Put & 27 & 72.60 & 74.70 & 22\\
 \hline
\end{tabular}
}
\captionsetup{width=1\textwidth}
\caption{Market quotes on 8 April 2016 at 14:55:00 for the forward, a call and a put option maturing 17 June 2016. For the forward, the bid and ask price quotes are the forward prices for entering a short or a long position, respectively. The data was extracted from Bloomberg.}
\label{table:BlooembergQuotes}
\end{table}

\section{The portfolio optimisation model}\label{sec:po}

For given initial wealth and quotes on cash, forward and the options, our aim is to find a portfolio with optimal net payoff at maturity. In general, the payoff will depend on the value of the underlying at maturity so the optimality will depend on our {\em risk preferences} concerning the uncertain payoffs. The optimality of a portfolio also depends on our financial position which may involve uncertain cash-flows at time $T$.


We will denote our initial wealth by $w\in\reals$ and assume that our financial position obligates us to pay $c$ units of cash at time $T$. The collection of all traded assets (cash, forward, options) is denoted by $J$. The cost of buying $x^j$ units of asset $j\in J$ is given by
\[
S^j_0(x^j) :=
\begin{cases}
s^j_a x^j & \text{if $x^j\ge 0$},\\
s^j_b x^j & \text{if $x^j\le 0$},
\end{cases}
\]
where $s^j_b\le s^j_a$ are the bid and ask prices of $j$. If $j$ is cash, we simply have $s^j_b=s^j_a=1$ while for the forward contract $s^j_b=s^j_a=0$. The finite quantities for the best quotes mean that there are upper and lower bounds $q^j_a$ and $q^j_b$, respectively, on the position $x^j$ one can take in asset $j$ at the best available quotes. For example, the quotes for the forward contract in Table~\ref{table:BlooembergQuotes} mean that $q^j_a=377$ while $q^j_b=-258$.

We will denote the payout of holding $x^j$ units of asset $j\in J$ by $P^j(x^j)$. The functions $P^j$ are given in Table~\ref{table:payoffs}. We model the value $X_T$ of the underlying at maturity as a {\em random variable} so that, in the case of forwards and the options, $P^j(x^j)$ will be random as well. We will assume that our financial before the trade obligates us to deliver a random amount $c$ of cash at maturity.

Modelling our risk preferences with expected utility, the portfolio optimization problem can be written as

\begin{equation}\label{alm}\tag{P}
\begin{aligned}
&\minimize\quad& &Ev(c-\sum_{j\in J}P^j(x^j))\quad\ovr\ x\in D\\
&\st\quad& & \sum_{j\in J}  S^j_0 (x^j) \leq w,
\end{aligned}
\end{equation}
where 
\[
D:=\prod_{j\in J}[q^j_b,q^j_a]
\]
is the set of feasible portfolios, $E$ denotes the expectation and $v(c):=-u(-c)$ with $u$ being the utility function. In the terminology of \cite{fs11}, $v:\reals\to\reals$ is a {\em loss function}. The argument of $v$ is the unhedged part of the claim $c$. Besides the available quantities, one could also include various margin requirements in the constraints.

It is clear that problem \eqref{alm} is highly subjective. Its optimum value and solutions depend on our
\begin{itemize}
\item
  financial position described by the initial cash $w$ and liability $c$,
\item
  views on the underlying $X_T$ described by the probabilistic model,
\item
  our risk preferences described by the loss function $v$.
\end{itemize}
The dependence will be studied numerically in the following sections. In pricing of contingent claims, the subjective factors will be reflected in the prices at which we are willing to trade the claims. The subjectivity is the driving force behind trading in practice but it is neglected e.g.\ by the traditional risk neutral pricing models.

Another important feature of \eqref{alm} is that it is a {\em convex optimization problem} as soon as the loss function $v$ is convex. The convexity simply means that we are risk averse. Convexity is crucial in numerical solution of \eqref{alm} as well as in the mathematical analysis of the indifference prices based on the optimum value of \eqref{alm}.



\section{Numerical portfolio optimization}

The first challenge in the numerical solution of problem~\eqref{alm} is that the objective is given in terms of an integral which, in general, does not allow for closed form expressions that could be treated by numerical optimization routines. However, in applications where the liability $c$ only depends on the the value of the underlying at maturity, the integral is one-dimensional which can be treated fairly easily with integration quadratures. This will be the case in the applications below where we study pricing and hedging of claims contingent on the underlying price at maturity. We will approximate the expectation by Gauss-Legendre quadrature which results in an objective given as a finite sum of convex functions of the portfolio vector $x$.

We will reformulate the budget constraint as two linear inequality constraints by writing the position in each asset as the sum of the long and short position. That is, $x^j=x^j_+-x^j_-$, where both $x^j_+$ and $x^j_-$ are constrained to be positive. This results in an inequality constrained convex optimization problem with the objective and constraints represented by smooth functions. The problem has 884 variables and 1769 constraints.

The resulting problem is solved with the interior-point solver of MOSEK~\cite{mosek} which is suitable for large-scale convex optimisation problems. To set up an instance of the optimization problem in MATLAB takes on average 11.20 seconds and its solution with MOSEK, 4.30 seconds on a PC with Intel(R) Core(TM) i5-4690 CPU @ 3.50GHz processor and 8.00 GB memory.

\subsection{Quotes, views and preferences}\label{sec:num}

We used quotes for S\&P500 index options with maturity 17 June 2016. The quotes were obtained from Bloomberg on 8 April 2016 at 2:55:00PM when the value of S\&P500 index was $2056.32$. The available quantities at the best quotes are given in terms of lot sizes which are 50 for forwards and 100 for options. The lending and borrowing rates are $0.0043$ and $0.03$, respectively, which correspond to the 1-month LIBOR rate and the borrowing rate of Yorkshire bank that offered the most generous rate at the time.

As a base case, we modelled the logarithm of the S\&P index at maturity with the Student t-distribution with the scale parameter $\sigma$ and degrees of freedom $\nu$ estimated from 25 years of historical daily data. The mean $\mu$ was set to zero. The effect of varying the parameters will be studied later on.

%

\begin{table}[h!]
  \centering
  \begin{tabular}{|c|c|c|}
\hline
    $\mu$ & $\sigma$ & $\nu$\\
\hline
    0.0000 &   0.0554 & 4.8355  \\
\hline
   \end{tabular}
  \captionsetup{width=1\textwidth}
  \caption{The parameters for the Student t-distribution used to model the index value at maturity.}
  \label{table:parameters}
\end{table}


As for the objective, we used the loss function
\[
v(c) = e^{\lambda c/w},
\]
where $w$ is the initial wealth and $\lambda>0$ is the risk aversion parameter. In other words, the risk preferences are described by exponential utility. It should be noted that, in general, the net position at maturity can take both positive as well as negative values which prevents the use of utility functions with constant relative risk aversion. The initial wealth $w$ used in the examples was $w=100,000$USD.


\subsection{The results}

\begin{figure}[H]
  \centering
  \includegraphics[width=1\textwidth,height=0.7\textwidth]{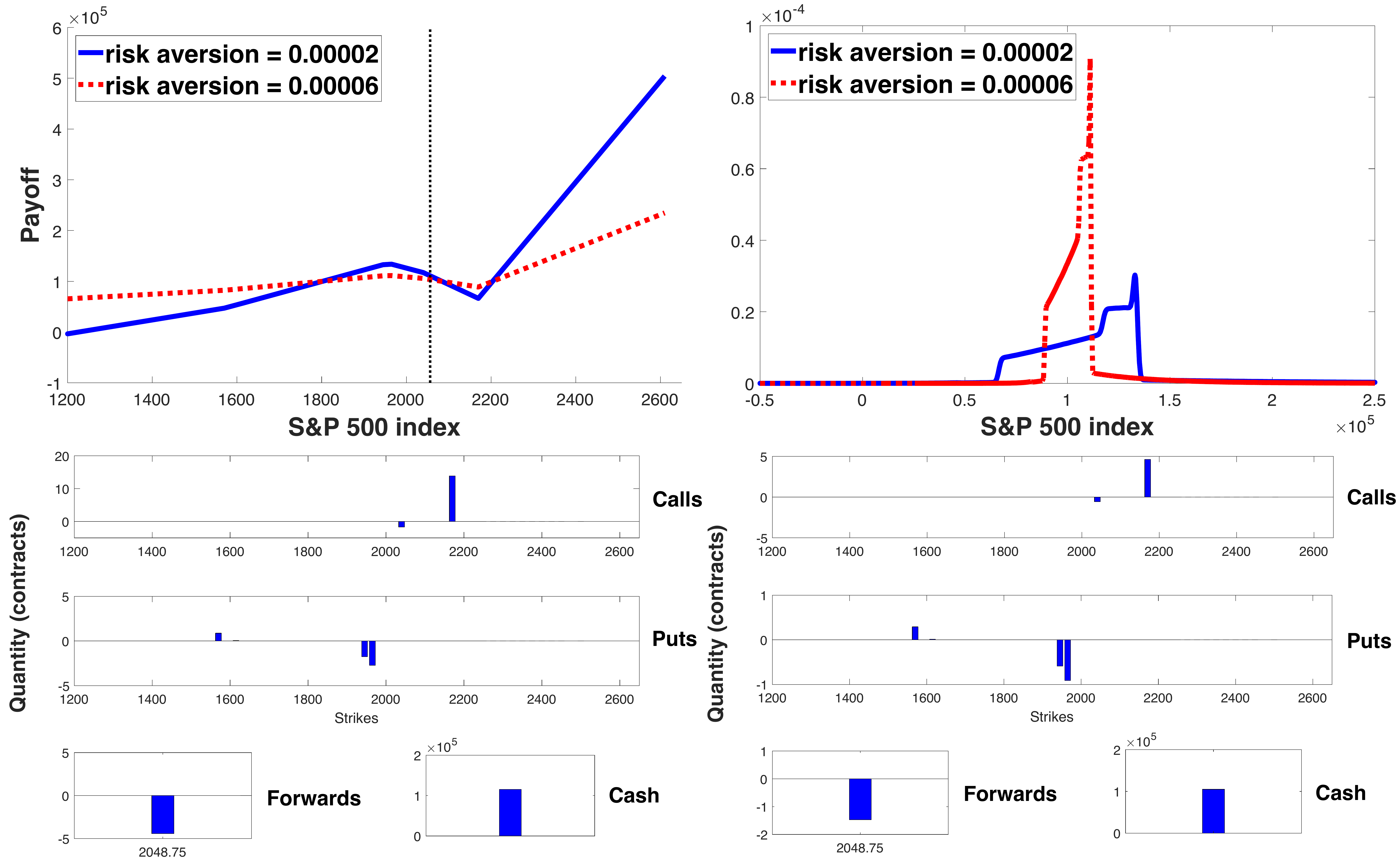}
\caption{The optimal portfolios obtained with risk aversions $\lambda=2$ and $\lambda=6$, respectively (bottom), the payoffs of the optimal portfolios as functions of the index at maturity (top-left) and the kernel-density estimates of the payoff distributions of the optimal portfolios (top-right)} \label{fig:figure5}  
\end{figure}

Figure~\ref{fig:figure5} illustrates the optimized portfolios obtained with two different risk aversions, $\lambda=2$ (blue line) and $\lambda=6$ (red line). The bottom panels represent the optimal portfolios with the bars corresponding to the optimal positions in the assets. The top left plots the corresponding payoffs as functions of the index at maturity and the top right plots the kernel density estimates (computed using 10,000,000 simulated values of the index at maturity) of the payoff distributions. As expected, higher risk aversion results in a payoff distribution with a thinner left tail. Increasing the risk aversion also results in reduced quantities in the optimal portfolio compared with the portfolio of a less risk averse agent.

\begin{figure}[H]
  \centering
    \includegraphics[width=1\textwidth,height=0.55\textwidth]{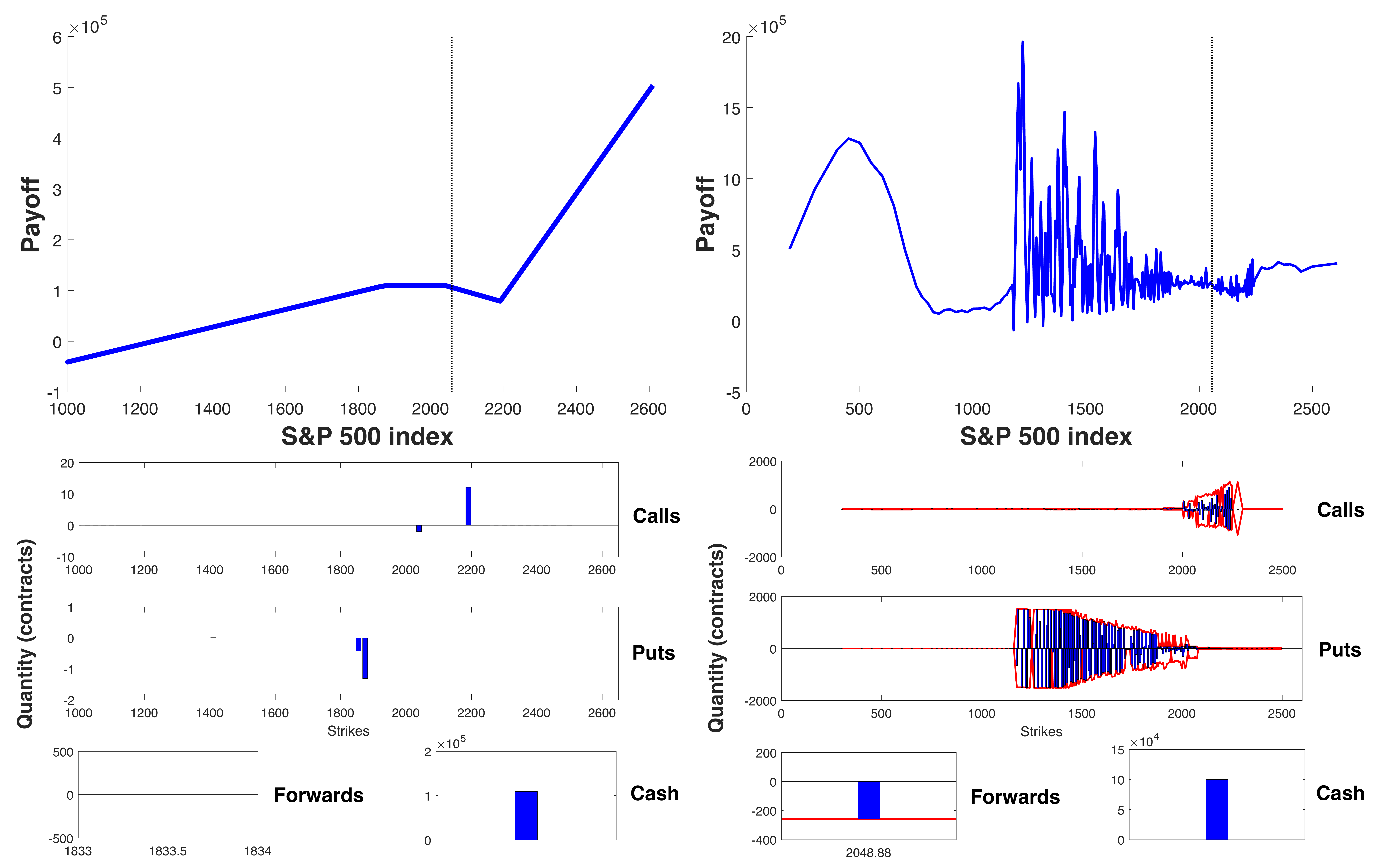}
    \caption{The payoffs and optimal portfolios when an additional 10\% transaction cost is added to all trades (left) and when the bid-ask spread is ignored by setting both bid- and ask-prices equal to the mid-price (right)} \label{fig:figure7}  
\end{figure}

An interesting feature of the optimal portfolios is that they are sparse in that our of more than 400 quoted options, the optimal portfolio has nonzero positions in less than 10 options. This is explained by the spreads between the quotes bid- and ask-prices. To illustrate this further, we repeated the optimization with risk aversion $\lambda=2$ by optimizing two variants of the problem. In the first one, we increased the bid-ask spread by adding a 10\% transaction cost on all trades and in the second, we set both the bid- and ask-prices equal to mid-prices. The results are illustrated in Figure~\ref{fig:figure7}. The addition of the transaction cost made the optimal portfolio only slightly sparser while removal of the bid-ask spread had a dramatic effect by giving a portfolio that takes large positions in almost all the quoted options. For many options, it was optimal to take maximal positions allowed by the available bid/ask quantities. 

\begin{figure}[H]
  \centering
  \includegraphics[width=0.65\textwidth,height=0.85\textwidth]{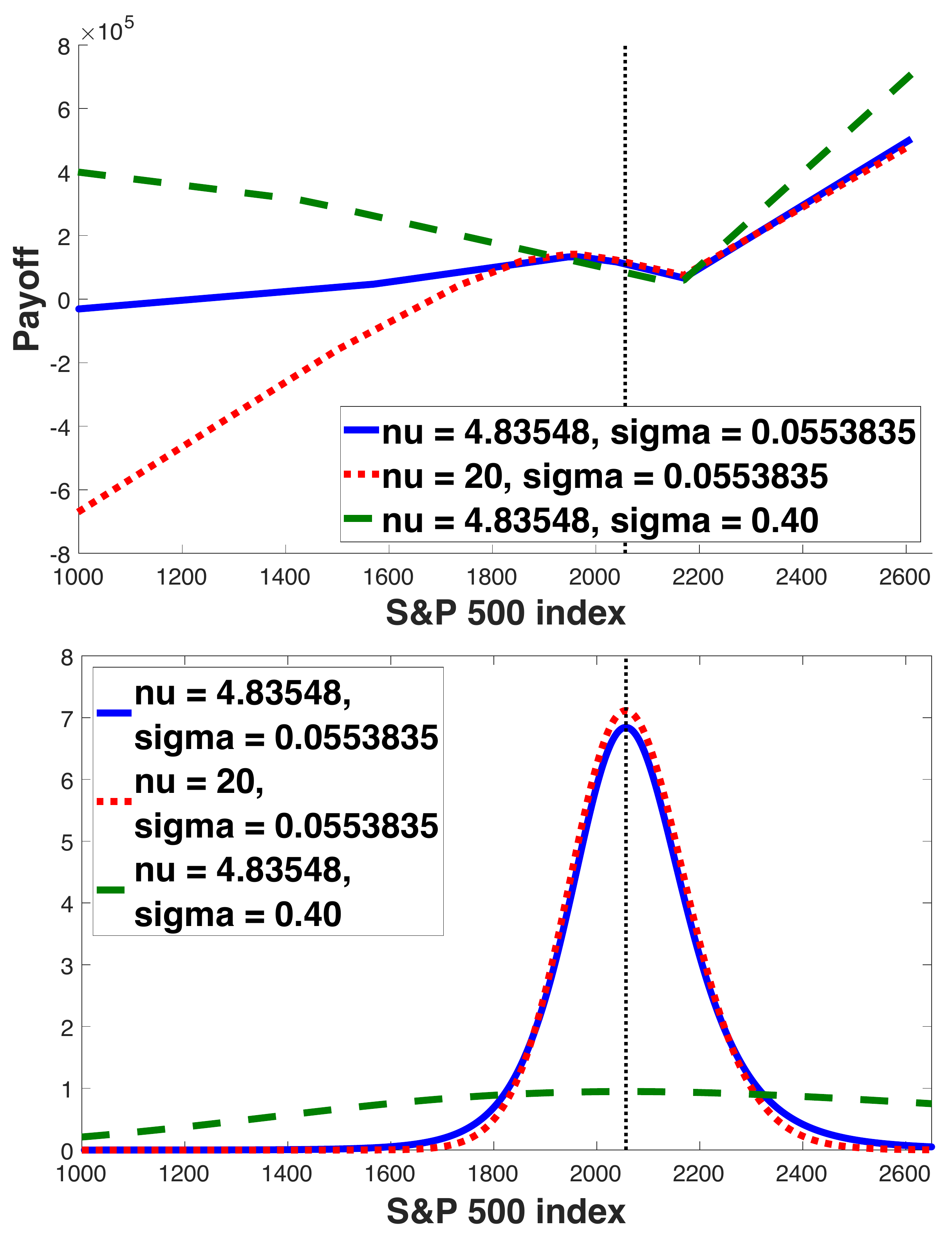}
  \caption{Distributions of the underlying (bottom) and optimal payoffs (top) in the base case (solid line), $\nu=20$ (dotted) and $\sigma=0.40$ (dashed). All other model parameters were unchanged.} \label{fig:figure6}  
\end{figure}

To study the effect of views on the optimal portfolio, we reoptimized the portfolio after changing the parameters of the underlying t-distribution. The risk aversion was kept at $\lambda=2$. Figure~\ref{fig:figure6} plots the payouts of the optimal portfolios in three cases. The first one is the base case already presented in Figure~\ref{fig:figure5}. The second if obtained by increasing the scale parameter $\sigma$ to $0.40$ and the third one by increasing the degrees of freedom $\nu$ to $20$. As expected, increasing $\sigma$ results in a portfolio that gives higher payouts further in the tails (a straddle) while $\nu=20$ gives essentially a Gaussian distribution with thinner tails so the optimal portfolio has higher payouts near the median at the expense of lower payoffs in the tails.

\begin{table}
  \centering
  \begin{tabular}{|l|c|}
    \hline
    Base case & -2.1499\\
    \hline
    $\sigma=0.40$ & -3.5121 \\
    \hline
    $\nu=20$ & -2.2339\\
    \hline
  \end{tabular}
  \caption{Logarithms of the objective values corresponding to the three different models of the underlying}
  \label{tab:erm}
\end{table}

\begin{figure}[H]
  \centering
    \includegraphics[width=1\textwidth,height=0.55\textwidth]{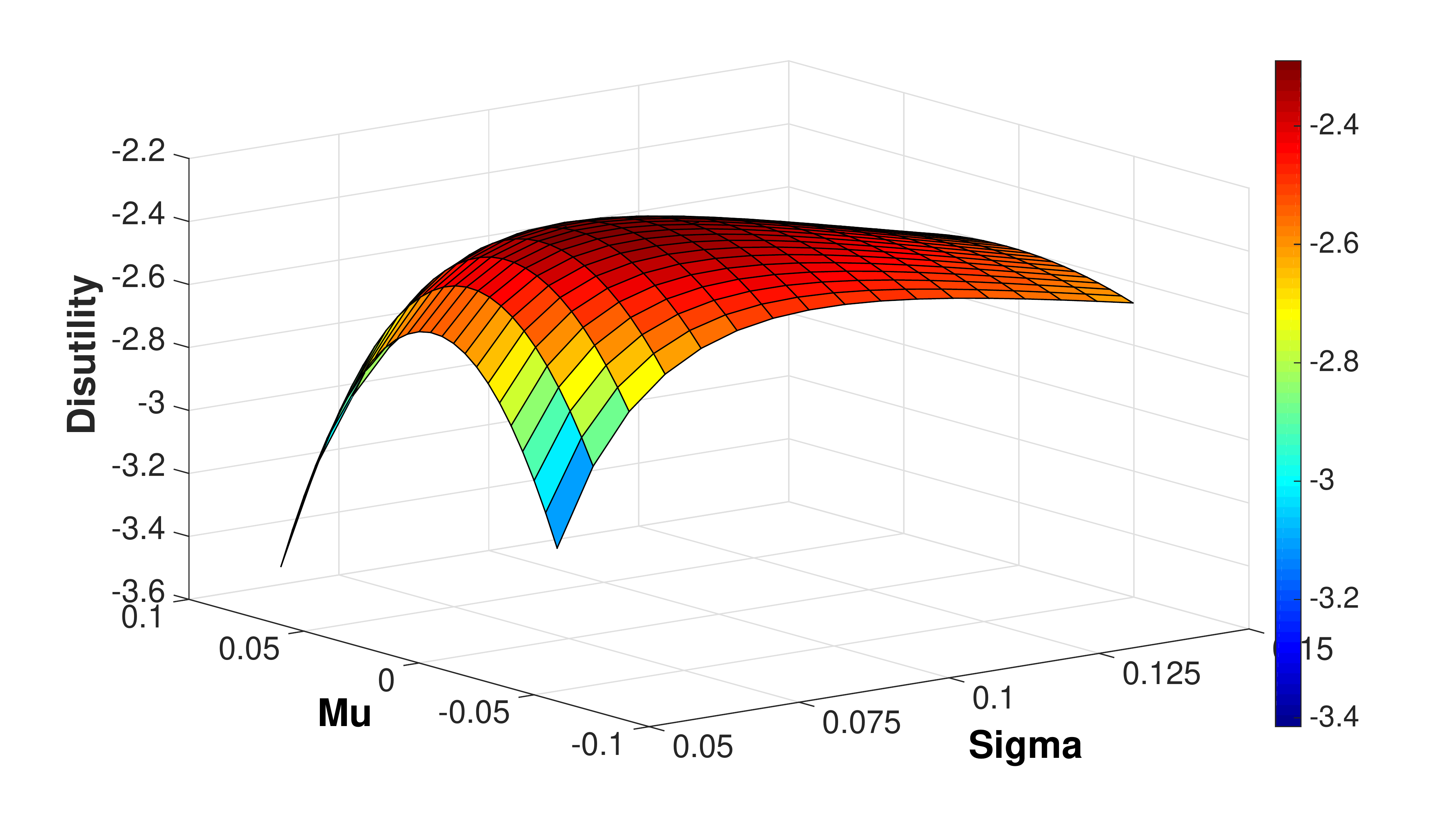}
  \caption{The entropic risk of the optimal portfolios as a function of the mean $\mu$ and volatility $\sigma$ when $\nu=\infty $ } \label{fig:3d}  
\end{figure}


The logarithms of the objective values obtained with the three models of the underlying in Figure~\ref{fig:figure6} are given in Table~\ref{tab:erm}. The logarithm of the expected exponential utility is known as the {\em entropic risk measure}; see e.g.\ \cite{fs11}. We see that the highest objective value is obtained with in the base case where the model parameters are estimated from historical data. An explanation of this could be that the option prices used in the model correspond to the market participants' views of the future behaviour of the underlying. If we use a model that is ``inconsistent'' with these prices, the option prices appear to offer profitable trading opportunities.


To explore this phenomenon more systematically, we repeated the optimization in the Gaussian case with $\nu=\infty$ and the mean $\mu$ and volatility $\sigma$ ranging over intervals. Figure~\ref{fig:3d} plots the corresponding logarithmic objective value, i.e.\ the entropic risk measure as a function of $\mu$ and $\sigma$. The risk seems to be concave as a function of ($\mu$,$\sigma$) with the maximum around $(\mu,\sigma)=(-0.05,0.08)$. The maximum value is $-2.289$. 

\section{Indifference pricing}

We will denote the optimum value of \eqref{alm} by
\[
\varphi(w,c):=\inf\{Ev(c-\sum_{j\in J}P^j(x^j))\,|\, x\in D,\ \sum_{j\in J} S^j_0(x^j) \leq w\}.
\]
For an agent with financial position $(\bar w,\bar c)$, the {\em indifference price} for selling a claim $c$ is given by
\[
\pi_s(\bar{w},\bar{c};c):=\inf\{w\,|\,\varphi(\bar{w}+w,\bar{c}+c)\leq\varphi(\bar{w},\bar{c})\}.
\]
This is the minimum price at which the agent could sell the claim $c$ without worsening her financial position as measured by the optimum value of \eqref{alm}. Analogously, the indifference price for buying $c$ is given by
\[
\pi_b(\bar{w},\bar{c};c):=\sup\{w\,|\,\varphi(\bar{w}-w,\bar{c}-c)\leq\varphi(\bar{w},\bar{c})\}.
\]
We have
\[
\pi_b(\bar w,\bar c;c)\le\pi_s(\bar w,\bar c;c)
\]
as soon as $\pi_s(\bar w,\bar c;0)=0$. Indeed, it is easily checked that the function $c\mapsto\pi_s(\bar w,\bar c;c)$ is convex so
\[
\pi_s(\bar w,\bar c;0)\le\frac{1}{2}\pi_s(\bar w,\bar c;c) + \frac{1}{2}\pi_s(\bar w,\bar c;-c)
\]
while $\pi_s(\bar w,\bar c;-c)=-\pi_b(\bar w,\bar c;c)$, by definition.

We will compare the indifference prices with the super- and subhedging costs defined for a claim $c$ by
\begin{align*} 
\pi_{\sup}(c)&:=\inf\{\sum_{j\in J}S^j_0(x^j)\,|\, x\in D,\ \sum_{j\in J} P^j(x^j)-c\ge 0\ P\text{-}a.s.\},\\
\pi_{\inf}(c)&:=\sup\{-\sum_{j\in J} S^j_0(x^j)\,|\, x \in D,\ \sum_{j\in J} P^j(x^j)+c \ge 0\ P\text{-}a.s.\}.
\end{align*}
The superhedging cost is the least cost of a superhedging portfolio while the subhedging cost is the greatest revenue one could get by entering position that superhedges the negative of $c$. Whereas the indifference prices of a claim depend on our financial position, views and risk preferences described by $(w,c)$, $P$ and $v$, respectively, the super- and subhedging costs are independent of such subjective factors. In complete markets, the sub- and superhedging costs are equal for all claims $c$ but, in general, the super- and subhedging costs are too wide apart to be considered as competitive quotes for a claim.

Recall that if $c:\reals_+\to\reals$ is the difference of convex functions, then its right-derivative is of bounded variation and we have 
\[
c(X_T) = c(0) + c'(0)X_T + \int_0^\infty(X_T-K)^+dc'(K).
\]
This might suggest that the payout $c$ could be replicated by a buy-and-hold portfolio of $c(0)$ units of a zero-coupon bond, $c'(0)$ units of the underlying and a continuum of call options weighted according to the Borel-measure associated with the BV function $c'$. Even if one could buy and sell options with arbitrary strikes, it is not quite realistic to trade a continuum of them. Nevertheless, assuming that quotes for all strikes exist, the replication cost of $c$ would become
\[
c(0)P_T + c'(0)X_0 + \int_0^\infty C(K)^adc'_+(K) - \int_0^\infty C(K)^bdc_-'(K),
\]
where $c'_+$ and $c'_-$ denote the positive and negative variations, respectively, of $c'$ and $C(K)^b$ and $C(K)^a$ denote the bid- and ask-prices of a call with strike $K$.

The above formulas could be used to design approximate replication strategies given the finite number of quotes in real markets. We will find out that the hedges optimized for indifference pricing look quite different from what the above replication approach would suggest. Instead of aiming for approximate replication, indifference pricing optimizes the portfolios to the given quotes, risk preferences and the given probabilistic description of the underlying.

\subsection{Numerical computation of indifference prices}

The definitions of the indifference prices involve the optimum value function $\varphi$ of problem \eqref{alm} which can rarely be evaluated exactly. The definitions still make sense, however, if we replace the optimum value by the best value we are able to find numerically. Besides the financial position, future views and risk preferences of an agent, the indifference prices then also depend on the agents' expertise in portfolio optimization. In computations below, we will replace $\varphi$ by the approximate value we find with the numerical techniques described in Section~\ref{sec:num}. The evaluation of the indifference prices then come down to a one-dimensional search over $w$. This can be done numerically by a line-search algorithm.

The computation of the super- and subhedging costs come down to solving linear programming problems where the constraints require the terminal position of the agent to be nonnegative in every scenario; see \cite{kkp5}. In the context of put and call options, the constraint can be written in terms of finitely many linear inequality constraints since we know that the net position will be linear between consecutive strike prices.

\subsection{Pricing exotic options}

We illustrate indifference pricing using the optimization model of Section~\ref{sec:po} in the pricing of three ``exotic'' options namely, a digital option with payoff
\[
c(X_T)=
\begin{cases}
  10,000 & \text{if $X_T\ge K$},\\
  0 & \text{if $X_T<K$}
\end{cases}
\]
a ``quadratic forward'' with $c(X_T)=|X_T-K|^2$ and a ``log-forward'' with $c(X_T)=100,000\ln(K/X_T)$, all with strike $K=2050$. Log-forwards have been used in the hedging of variance swaps; see e.g.\ \cite{cm98}. To compare with a simper option, we also price a European call option with the same strike. To make the last case nontrivial, we remove the call from the set of hedging instruments.

We compute the indifference selling prices assuming that $\bar w=100,000$ and $\bar c=0$, that is, assuming the agent has initial position consisting only of 100,000 units of cash. The indifference prices together with the super- and subhedging costs are given in Table~\ref{table:indiffPrices}. Superhedging is imposed on the interval [100,5000]. Clearly, superhedging the quadratic and log-forwards with the given hedging instruments against all positive values of $X_T$ is impossible. The numbers reported in Table~\ref{table:indiffPrices} are the costs of hedging over the interval [100,5000].

\begin{table}[h!]
  \centering
  \resizebox{1\textwidth}{!}{  
    \begin{tabular}{|l|c|c|c|c|} 
      \hline
      Claim & subhedging& buying price & selling price& superhedging\\ 
      \hline
      call & 51.2333 & 51.7338 & 51.7399 & 53.0483 \\ 
      digital call & 5280.00& 6082.35& 6160.65 &6885.71 \\ 
      quadratic forward & 20383.68& 20979.84&22044.92&  24542.01\\ 
      log-forward & 322.28& 358.49 &  404.67 &499.69\\ 
      \hline
    \end{tabular}
  }
  \captionsetup{width=1\textwidth}
  \caption{Indifference prices, together with super- and subhedging costs.}
  \label{table:indiffPrices}
\end{table}

\begin{figure}[H]
  \centering
  \includegraphics[width=1\textwidth,height=0.55\textwidth]{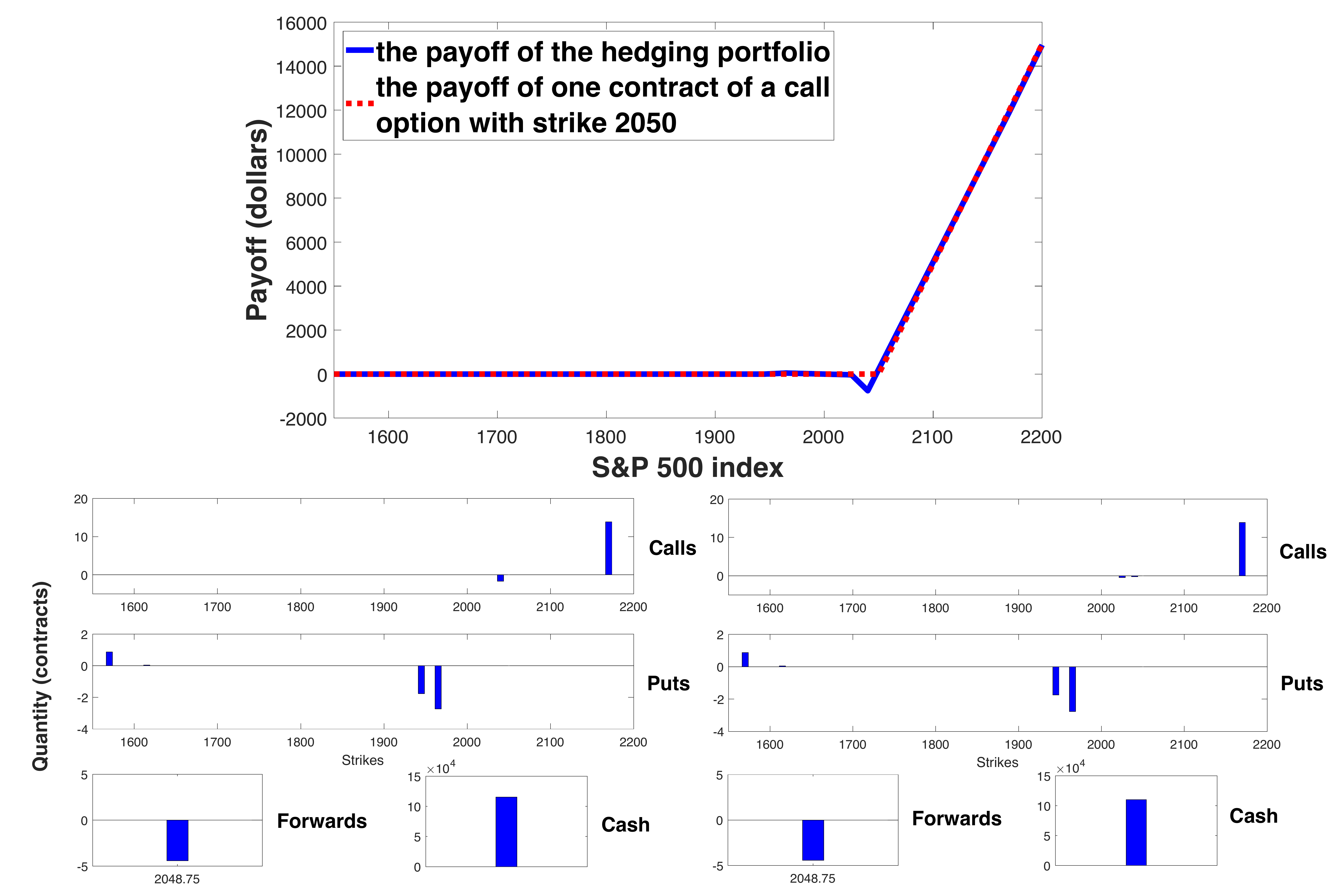}
  \caption{Optimal portfolios before (bottom left) and after (bottom right) the sale of a call option. The top panel gives the payoff of the hedging portfolio (solid line) together with the payoff of the claim being priced (dotted line).}
  \label{fig:sellCall}  
\end{figure}

\begin{figure}[H]
  \centering
  \includegraphics[width=1\textwidth,height=0.55\textwidth]{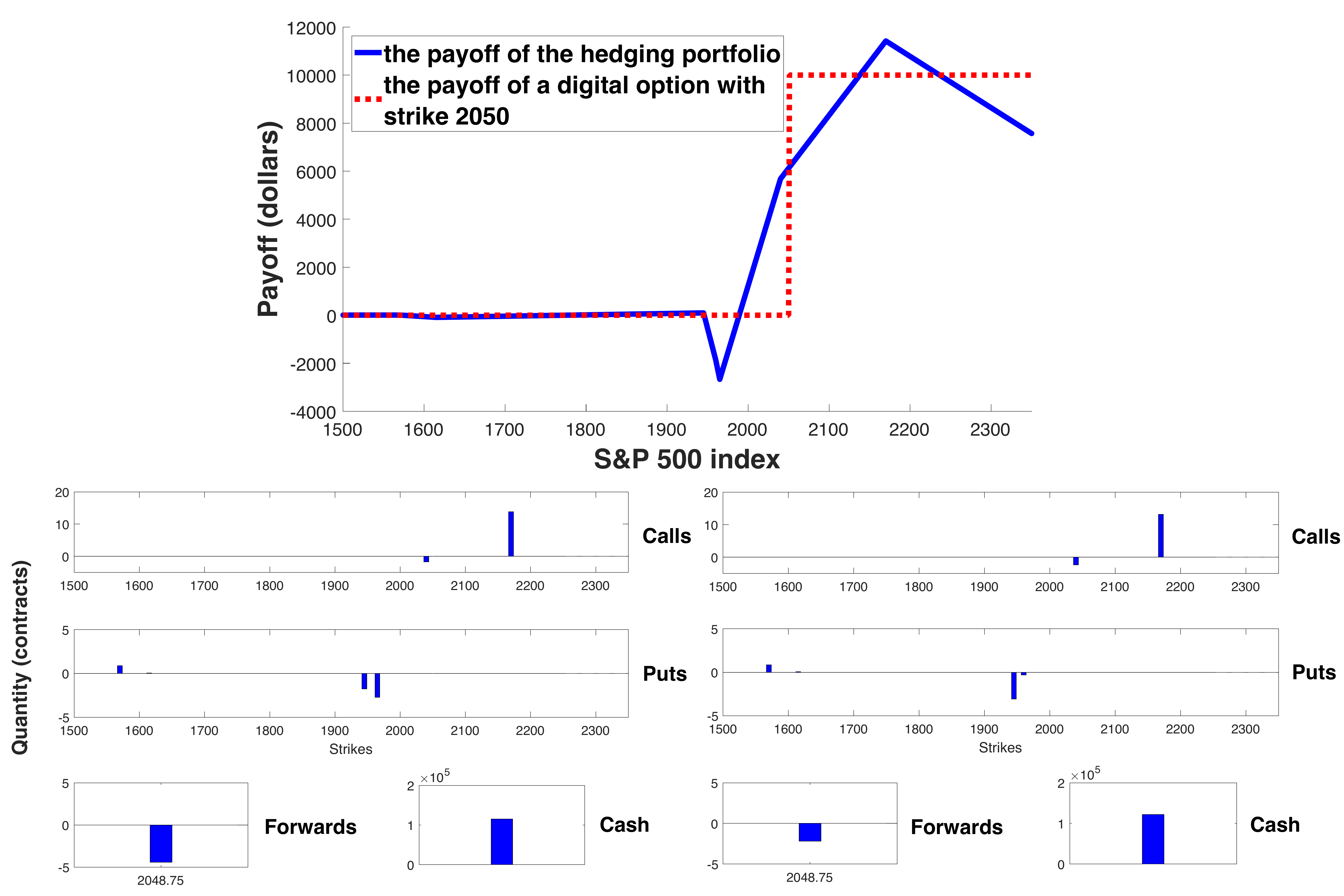}
  \caption{Optimal portfolios before (bottom left) and after (bottom right) the sale of a digital option. The top panel gives the payoff of the hedging portfolio (solid line) together with the payoff of the claim being priced (dotted line).}
  \label{fig:sellDigital}  
\end{figure}

\begin{figure}[H]
  \centering
  \includegraphics[width=1\textwidth,height=0.55\textwidth]{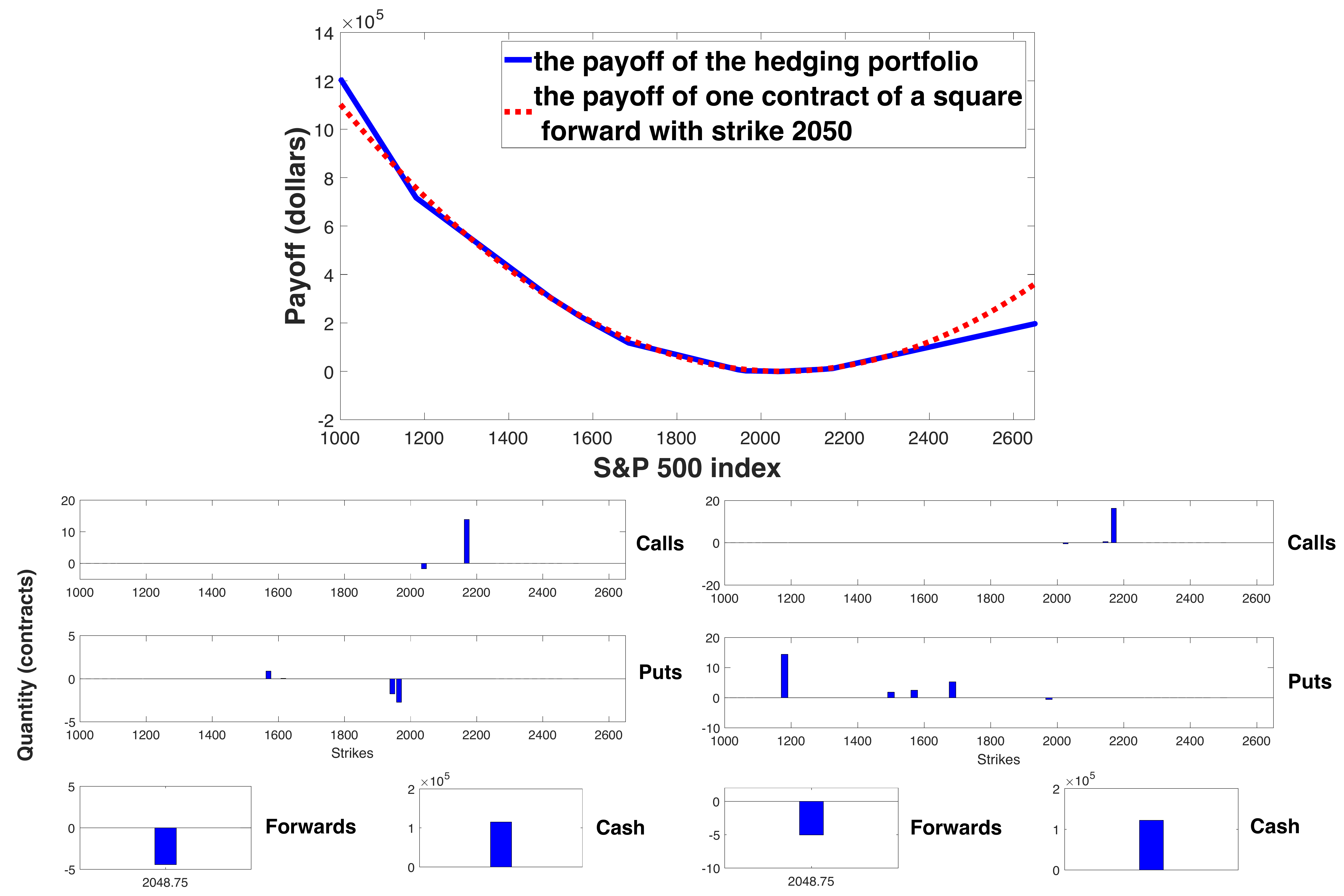}
  \caption{Optimal portfolios before (bottom left) and after (bottom right) the sale of a quadratic forward. The top panel gives the payoff of the hedging portfolio (solid line) together with the payoff of the claim being priced (dotted line).}
  \label{fig:sellParabola}  
\end{figure}

\begin{figure}[H]
  \centering
  \includegraphics[width=1\textwidth,height=0.55\textwidth]{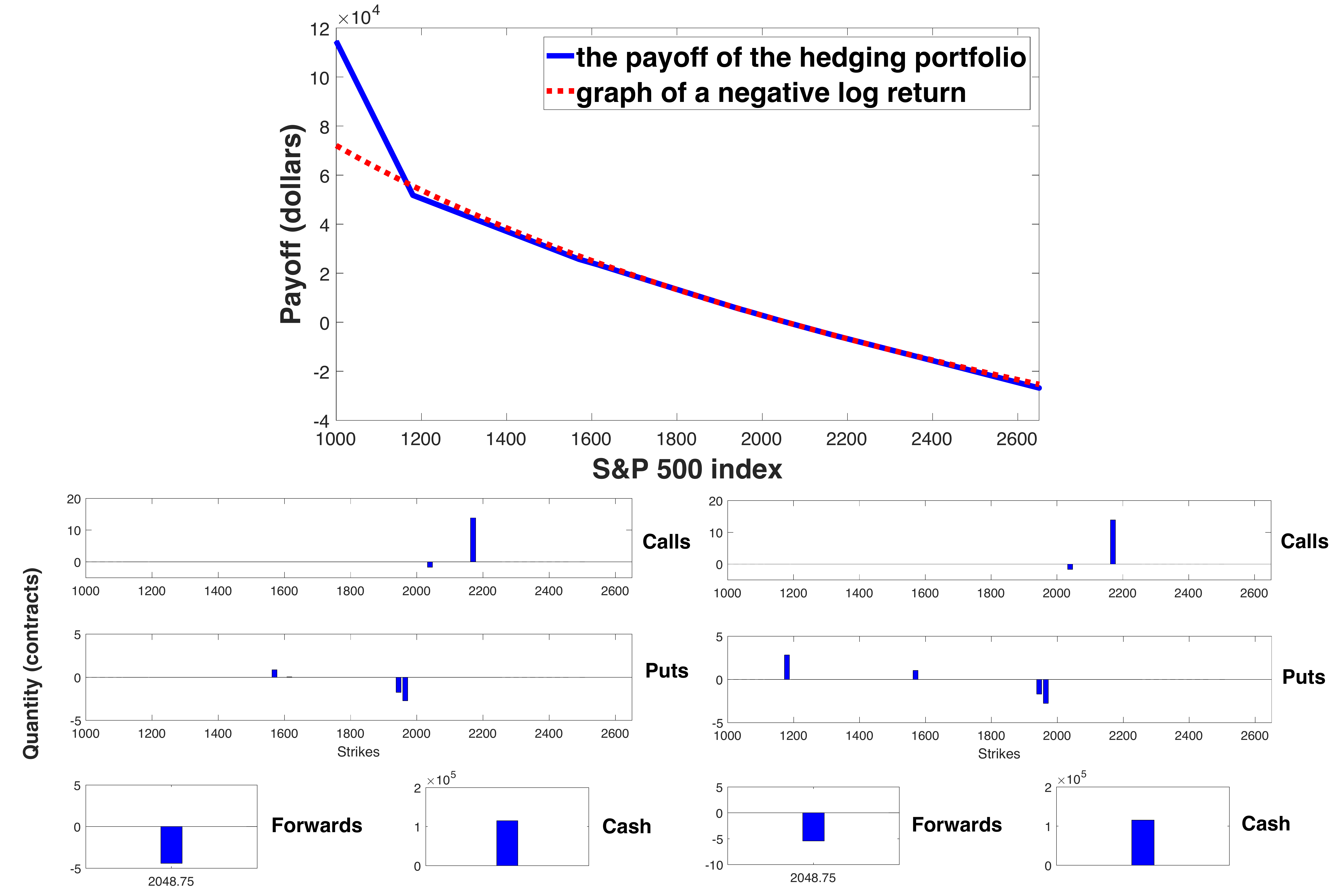}
  \caption{Optimal portfolios before (bottom left) and after (bottom right) the sale of a log-forward. The top panel gives the payoff of the hedging portfolio (solid line) together with the payoff of the claim being priced (dotted line).} 
  \label{fig:sellLogReturn}  
\end{figure}

Figures~\ref{fig:sellCall}--\ref{fig:sellLogReturn} illustrate the corresponding hedging strategies. Each figure gives the optimal portfolio before and after selling the option together with the payout of the``hedging portfolio'' as a function of the underlying at maturity. The {\em hedging portfolio} is defined as the difference $x-\bar x$, where $\bar x$ and $x$ are the optimal portfolios before and after the sale of the option.

\subsection{Sensitivities}

\begin{figure}[H]
  \centering
    \includegraphics[width=1\textwidth,height=0.55\textwidth]{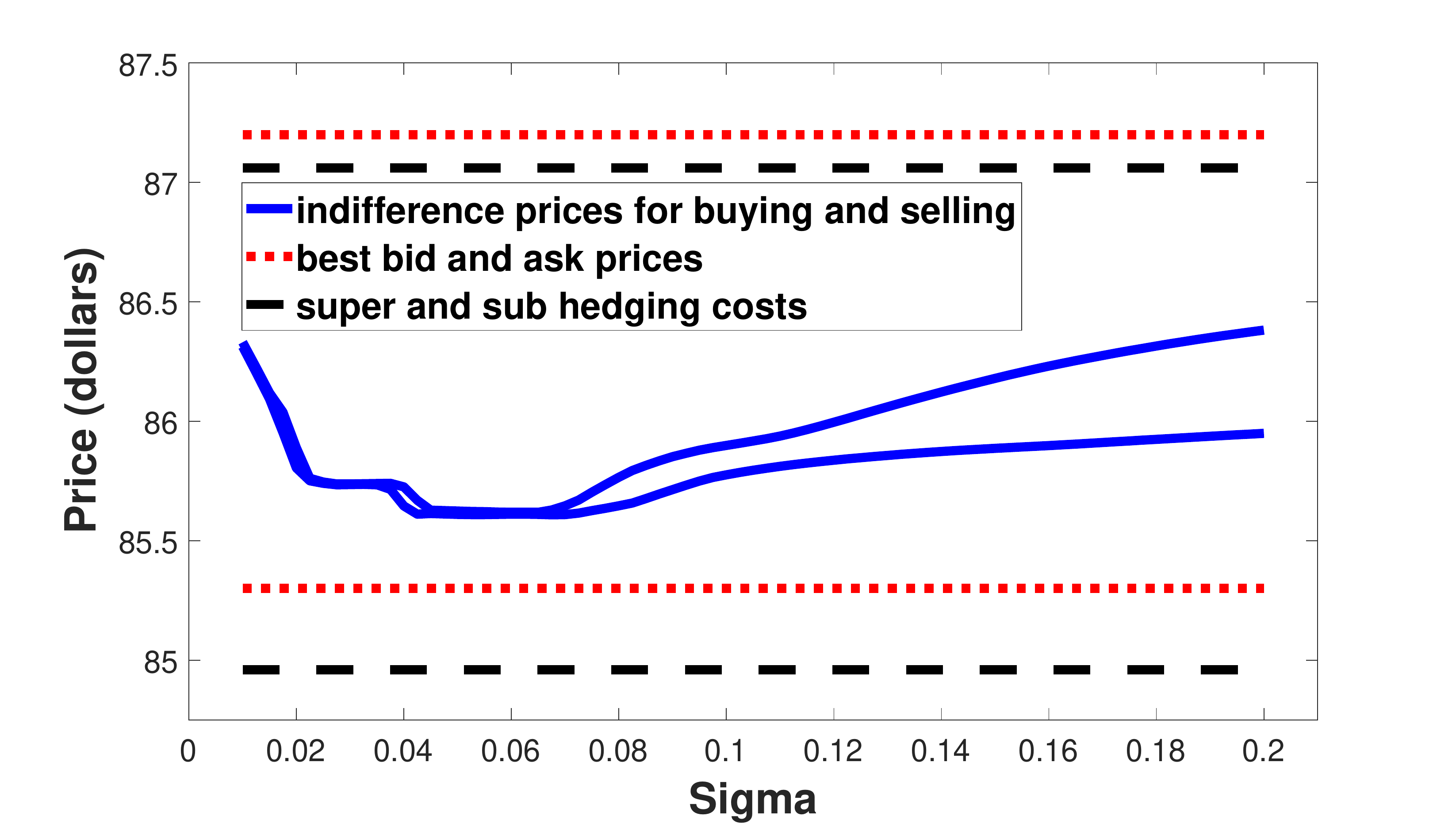}
  \caption{Indifference prices as functions of volatility. The dotted lines give the best bid and ask quotes while the dashed lines give the super- and subhedging costs.} \label{fig:difSigma}  
\end{figure}

This section studies the sensitivities of the indifference prices with respect to some of the model parameters. Figure~\ref{fig:difSigma} plots indifference prices of a call option with strike 2000 as functions of the ``volatility'' $\sigma$ (Since we model the underlying with the t-distributions, the variance of the log-price is $\sigma^2\nu/(\nu-2)$). Again, we have removed the call being priced from the set of hedging instruments when computing the prices. Instead of being monotone, the indifference prices achieve their minimums when $\sigma$ is close to its historical estimate of $0.0554$. The implied volatility computed with the classical Black--Scholes model from the mid-quote of the call is 0.1478.

\begin{figure}[H]
  \centering
    \includegraphics[width=1\textwidth,height=0.50\textwidth]{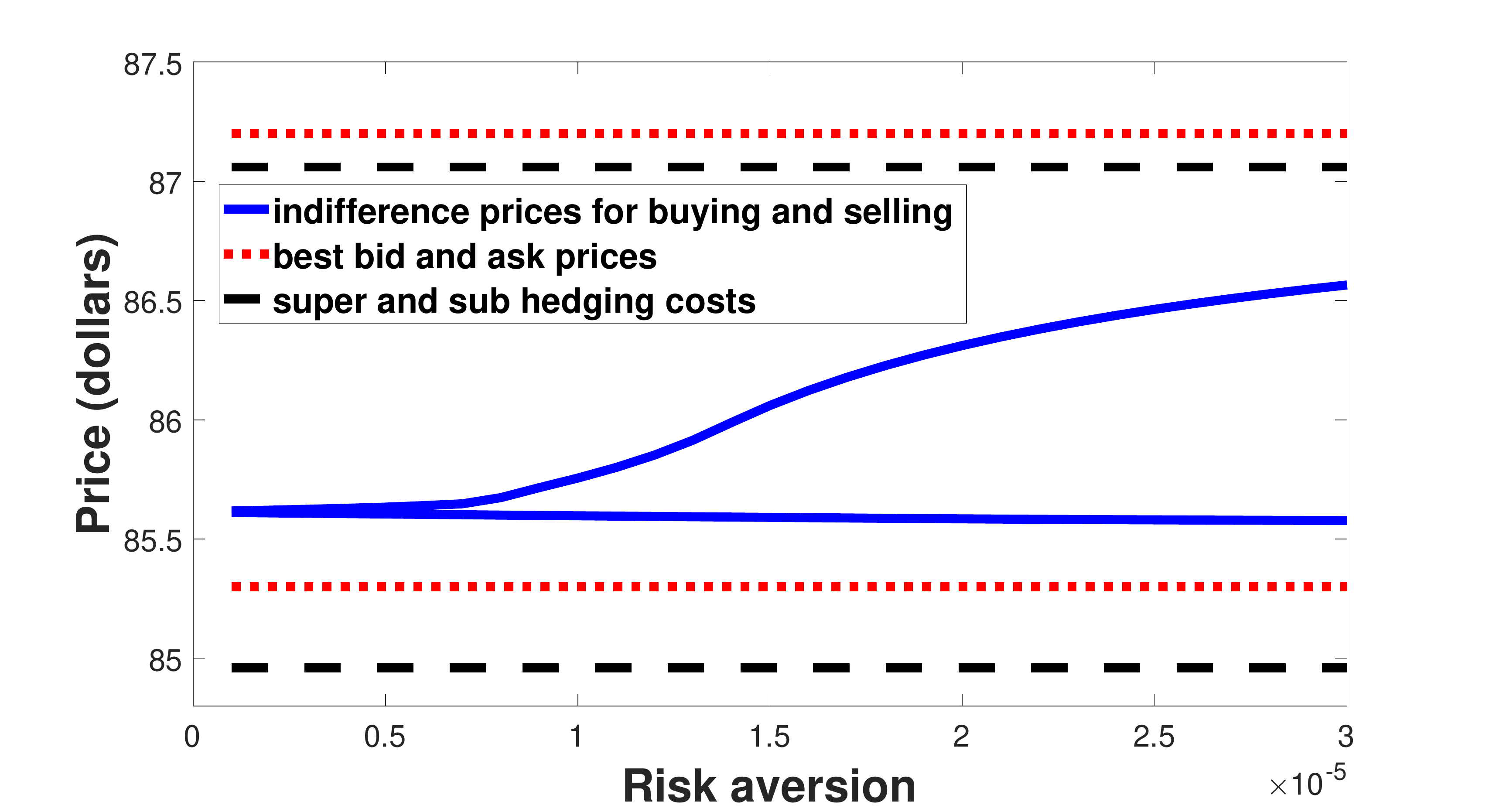}
  \caption{Indifference prices of a call option with strike 2000 as functions of risk aversions.} \label{fig:12}
\end{figure}

Figure~\ref{fig:12} plots the indifference prices as functions of the risk aversion. As the risk aversion increases, the gap between the indifference prices widens. The indifference price for selling a call option is more sensitive to the risk aversion. This seems quite natural as shorting a call results in unbounded downside risk unless the call is superhedged. 

\begin{figure}[H]
  \centering
    \includegraphics[width=1\textwidth,height=0.50\textwidth]{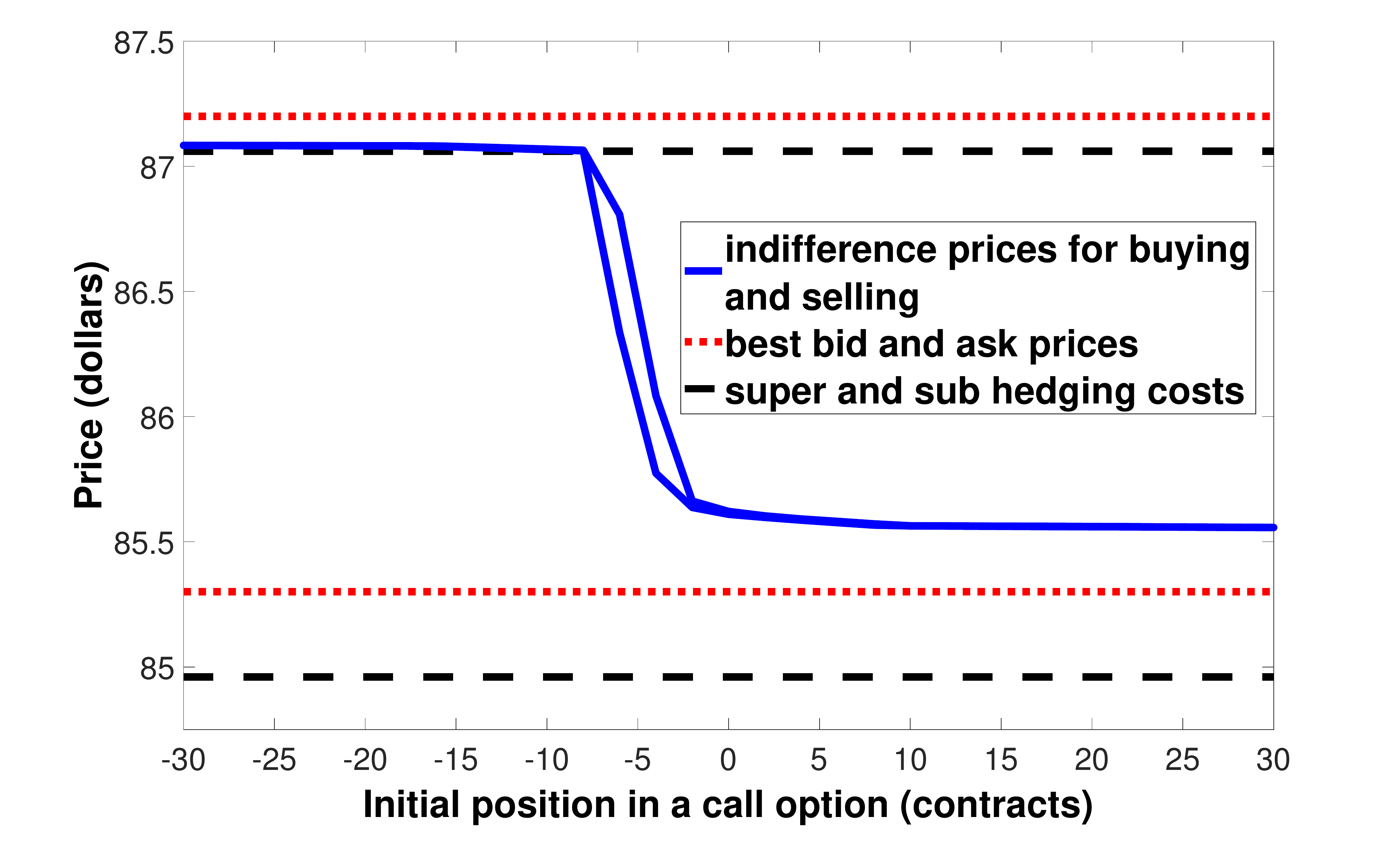}
  \caption{Indifference prices of a call option with strike 2000 as functions of initial position in the same call} \label{fig:ip}
\end{figure}

Figure~\ref{fig:ip} illustrates the dependence of the indifference prices on an agent's initial position. While in earlier cases, the agent's initial position was assumed to consist only of cash, in this case, we consider an agent with both cash and call options of the same type as the one being priced. Figure~\ref{fig:ip} plots the indifference prices as functions of the number of call options the agent holds before the trade. As one might expect, an agent who already has exposure to the option would assign a higher price to the option. A seller would increase her exposure to the option payout while for a buyer, the option would be a natural hedge and thus worth paying a higher price for.

\begin{figure}[H]
  \centering
  \includegraphics[width=1\textwidth,height=0.40\textwidth]{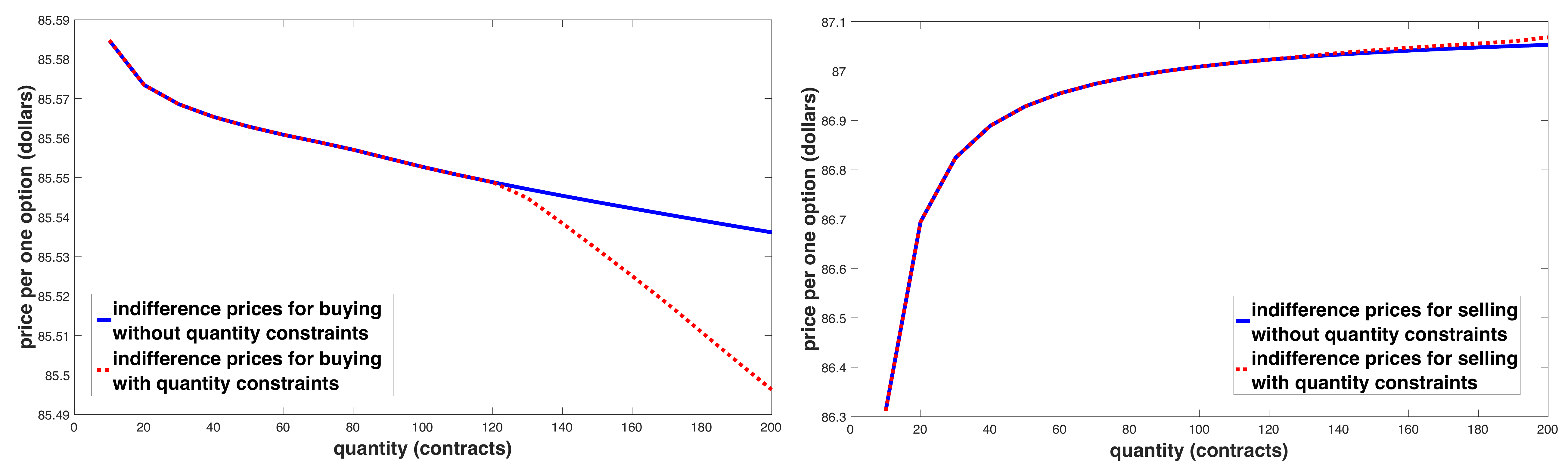}
  \caption{Indifference prices of a call option per unit as a function of the quantity traded. Buying price on the left and selling price on the right. The solid line gives the prices when quantity constraints are ignored} \label{fig:21}
\end{figure}

To illustrate the nonlinearity of the indifference prices as functions of the claim, we computed the prices for different multiples $M$ of the call. Figure~\ref{fig:21} plots the indifference prices per option as functions of the multiplier $M$. The figure plots the indifference prices also in a market model where the best quotes are assumed to come with unlimited quantities. As the multiplier $M$ increases, the quantity constraints become binding thus worsening the prices.

\section{Further developments}

The developed indifference pricing framework should be taken merely as an illustration of the computational techniques that are available for portfolio optimization. The presented model could be extended in various ways in practice. For example, it would be straightforward to include margin requirements as portfolio constraints in the model, as long as the requirements are given as explicit convex constraints on the portfolio. One could also study options with different maturities by including the relevant maturities in the underlying probabilistic model. Such a multiperiod model, could also incorporate dynamic trading strategies of the underlying and cash.

\bibliographystyle{plain}
\bibliography{put-call}

\end{document}